%% file: main.tex
\documentclass[sigconf]{acmart}
\AtBeginDocument{%
  }

\copyrightyear{2026}
\acmYear{2026}
\setcopyright{cc}
\setcctype{by}
\acmConference[CHI '26]{Proceedings of the 2026 CHI Conference on Human Factors in Computing Systems}{April 13--17, 2026}{Barcelona, Spain}
\acmBooktitle{Proceedings of the 2026 CHI Conference on Human Factors in Computing Systems (CHI '26), April 13--17, 2026, Barcelona, Spain}
\acmPrice{}
\acmDOI{10.1145/3772318.3790419}
\acmISBN{979-8-4007-2278-3/2026/04}

\begin{document}

\title{Understanding How Accessibility Practices Impact Teamwork in Mixed-Ability Teams that Collaborate Virtually}


\author{Crescentia Jung}
\email{cj382@cornell.edu}
\affiliation{%
  \institution{Cornell University}
  \city{New York}
  \state{NY}
  \country{USA}
}
\author{Kexin Cheng}
\email{kc2248@cornell.edu}
\affiliation{%
  \institution{Cornell Tech}
  \city{New York}
  \state{NY}
  \country{USA}
}
\author{Sharon Heung}
\email{ssh247@cornell.edu}
\affiliation{%
  \institution{Cornell Tech}
  \city{New York}
  \state{NY}
  \country{USA}
}
\author{Malte F. Jung}
\email{mfj28@cornell.edu}
\affiliation{%
  \institution{Cornell University}
  \city{Ithaca}
  \state{NY}
  \country{USA}
}
\author{Shiri Azenkot}
\email{sa933@cornell.edu}
\affiliation{%
  \institution{Cornell Tech}
  \city{New York}
  \state{NY}
  \country{USA}
}

\renewcommand{\shortauthors}{Jung et al.}

\begin{abstract}
\input{abstract}
\end{abstract}

\ccsdesc[500]{Human-centered computing~Empirical studies in HCI}

\keywords{accessibility, mixed-ability, teamwork, virtual collaboration}


\newcommand{\update}[1]{\textcolor{black}{#1}}

\maketitle
\input{final}

\bibliographystyle{ACM-Reference-Format}
\bibliography{references}

\section{Semi-Structured Interview Protocol}
Below is our semi-structured interview protocol. We asked probing and follow up questions beyond the questions listed here. 
\subsection{Participants and Their Teams’ Backgrounds}
\subsubsection{Demographics and Participant’s Background}
\begin{enumerate}
    \item What is your age?
    \item How do you identify in terms of ethnicity?
    \item How do you identify in terms of gender?
    \item What is your occupation?
    \begin{enumerate}
        \item What are your responsibilities?
        \item What is your workplace background?
    \end{enumerate}
    \item What is your education level?
    \item Do you identify as having a disability?
    \begin{enumerate}
        \item If so, please describe your disability.
        \item If so, what assistive technologies do you typically use?
    \end{enumerate}
\end{enumerate}
\subsubsection{Teamwork Activities and Background}
\begin{enumerate}
    \item Can you tell me about the last virtual meeting you had with your mixed-ability team?
    \item Can you tell me more about how you and your team use virtual meetings?
    \begin{enumerate}
        \item Can you describe what the virtual meetings are for?
    \end{enumerate}
    \item What happens before and after virtual meetings?
    \item How frequently do you have virtual meetings with your team?
    \item Which video conferencing platforms do you use for virtual meetings?
    \item How many people are usually in these virtual meetings?
    \item Could you tell me more about your team members?
    \begin{enumerate}
        \item Are you aware of their disability status?
        \item What is their role in the team?
    \end{enumerate}
\end{enumerate}
\subsubsection{Accommodations and Expectations}
Questions for disabled participants:
\begin{enumerate}
    \item Do you use any assistive technologies to make virtual meetings more accessible?
    \item Are there any accommodations, rules, or expectations that you or your team members practice to make meetings more inclusive and accessible?
    \begin{enumerate}
        \item Are they explicitly communicated to the meeting attendees?
        \item Are there any implicit norms that you’ve observed?
        \item What is challenging about maintaining accommodations, rules, or expectations? How did you overcome that?
    \end{enumerate}
    \item How do you communicate your accommodation needs to your team members?
    \begin{enumerate}
        \item Do you request these in advance or on the spot?
        \item Was there ever a time when your accommodation needs were not met? How did you overcome the challenge?
    \end{enumerate}
    \item Have you seen methods for gently or explicitly reminding attendees of accommodations?
    \item Do you experience any other challenges or access barriers in virtual meetings? If so, what are your experiences?
\end{enumerate}

Questions for non-disabled participants:
\begin{enumerate}
    \item Are there any accommodations, rules, or expectations that you or your team members practice to make meetings more inclusive and accessible?
    \begin{enumerate}
        \item Are they explicitly communicated to the meeting attendees?
        \item Are there any implicit norms that you’ve observed?
        \item What is challenging about maintaining accommodations, rules, or expectations? How did you overcome that?
    \end{enumerate}
    \item Have you seen methods for gently or explicitly reminding attendees of accommodations?
\end{enumerate}
\subsection{Teamwork}
\subsubsection{Camaraderie}
\begin{enumerate}
    \item You mentioned that you work with [team member]. Can you tell me more about your relationship with [team member]?
    \item When did you first meet [team member]?
    \item How often do you interact with [team member]?
    \item Do you feel connected with [team member]? Why or why not?
    \begin{enumerate}
        \item Can you give an example of a time when you felt particularly connected to [team member]?
    \end{enumerate}
    \item How much do you feel that you trust [team member]?
    \begin{enumerate}
        \item Are there any recent instances that affected your level of trust with [team member]?
    \end{enumerate}
    \item How personal are your interactions with [team member]?
    \begin{enumerate}
        \item Can you give an example of a meaningful interaction you've had recently with [team member]?
    \end{enumerate}
\end{enumerate}

\subsubsection{Productivity}
\begin{enumerate}
    \item How do you feel about your team’s productivity?
    \begin{enumerate}
        \item How do you feel that disability and accommodations affect productivity?
    \end{enumerate}
    \item How do you perceive your productivity when collaborating with your team members?
    \begin{enumerate}
        \item Can you describe your experiences working with [team member]?
    \end{enumerate}
    \item How do you collaborate with your team members to complete tasks?
    \begin{enumerate}
        \item What tools do you use to communicate?
        \item What tools do you use for collaborative tasks?
        \item You mentioned you use [tools] to collaborate. Are these effective in getting the tasks done? Why or why not?
        \item Can you provide an example of a recent successful collaboration experience?
        \item Can you provide an example of a recent unsuccessful collaboration experience?
        \begin{enumerate}
            \item How did you overcome that challenge?
        \end{enumerate}
        \item How does this affect your interactions during virtual meetings?
    \end{enumerate}
    \item How effective do you feel in accomplishing your work with team members?
    \begin{enumerate}
        \item Can you recall any recent occasions when you felt that it was difficult to accomplish work effectively when collaborating virtually with your team members?
        \begin{enumerate}
            \item Could you describe a situation or experience where you encountered difficulties in getting work done as effectively as you'd like? How did you overcome that challenge?
        \end{enumerate}
    \end{enumerate}
    \item Is there anything that would improve your productivity when working with your team members?
\end{enumerate}

\subsubsection{Participation}
\begin{enumerate}
    \item What’s your perception of the level of participation among your team members?
    \begin{enumerate}
        \item For yourself?
        \item What’s your perception of [team member]’s participation? Why?
    \end{enumerate}
    \item How comfortable are you in participating in meetings?
    \begin{enumerate}
        \item Can you recall a recent instance when you felt uncomfortable participating in meetings? If so, how did you overcome that challenge?
    \end{enumerate}
    \item How comfortable are you with expressing your opinions during meetings?
    \begin{enumerate}
        \item Can you recall a recent instance when you felt uncomfortable sharing your opinion in meetings? If so, how did you overcome that challenge? 
    \end{enumerate}
    \item How do you feel your participation changes over the different types of meetings you have?
\end{enumerate}
\subsection{Reflection}
\begin{enumerate}
    \item What do you think makes for “good” teamwork with your mixed-ability team?
    \item How do disability, accessibility, and accommodations affect your mixed-ability team?
    \item Is there anything you’d like to share or questions you think I missed before we end the interview?
\end{enumerate}
\section{Examples of Codes}

\begin{table}[H]
\caption{Examples of quotes and codes applied to the quote.}
\label{tab:codes-examples}
\small
\setlength{\tabcolsep}{6pt}
\renewcommand{\arraystretch}{1.15}
\begin{tabular}{@{}p{0.40\columnwidth} p{0.52\columnwidth}@{}}
\toprule
\textbf{Quote} & \textbf{Applied Codes} \\
\midrule

“The way we meet is like a family; we check in and always check on each other to find out if people are okay and have the accommodations they need.” (D10)
& consistent accessibility check-ins, building rapport through routines \\
    \addlinespace

“We usually have a sheet that everybody writes specifically what accommodations they may need… I think that just helps build rapport within the team as well as destigmatize.” (D6) 
& 
shared accessibility document for building rapport, visibility of access needs, importance of destigmatizing accommodations \\
    \addlinespace

“One of my family members was going through chemotherapy, and so, I would put that in and be like, ‘I'm gonna be gone like helping out.’” (D11) 
& sharing non-accessibility related needs, shared accessibility document for coordination \\
    \addlinespace

“I don’t think I would think to ask, ‘Hey, can we do this?’ But someone else initiating it would normalize it. Like, oh, that person asked. It’s a thing, a norm, a cultural shift.” (D5)
& 
accessibility as an evolving norm, shifts in team culture for inclusivity, initiation in access norm reduces stigma  \\

\bottomrule
\end{tabular}
\end{table}

\color{black}

\end{document}

%% file: abstract.tex
Virtual collaboration has transformed how people in mixed-ability teams, composed of disabled and non-disabled people, work together by offering greater flexibility. In these settings, accessibility practices, such as accommodations and inclusive norms, are essential for providing access to disabled people. However, we do not yet know how these practices shape broader facets of teamwork, such as productivity, participation, and camaraderie. To address this gap, we interviewed 18 participants (12 disabled, 6 non-disabled) who are part of mixed-ability teams. We found that beyond providing access, accessibility practices shaped how all participants coordinated tasks, sustained rapport, and negotiated responsibilities. Accessibility practices also introduced camaraderie challenges, such as balancing empathy and accountability. Non-disabled participants described allyship as a learning process and skill shaped by their disabled team members and team culture. Based on our findings, we present \update{recommendations for team practices and design opportunities} for virtual collaboration tools that reframe accessibility practices as a foundation for strong teamwork.

%% file: final.tex
\section{Introduction}
The COVID-19 pandemic prompted a global shift to remote work, accelerating the use of virtual collaboration tools and transforming how teams communicate and work together \cite{noauthor_virtual_nodate}. In these \textit{virtual settings}, team members collaborate in remote environments using digital tools. These include synchronous interactions, such as meetings on platforms like Zoom, and asynchronous collaboration through tools like shared document editing. Virtual collaboration has become essential for teamwork, offering increased flexibility for team members. For many people with disabilities, this shift can reduce physical barriers to participation and involve tools that often provide accessibility features, such as closed captioning, that support collaboration \cite{murphy_remote_2021, shew2020let}. 

Researchers have examined the accessibility of virtual collaboration, particularly in the context of \textit{mixed-ability teams} consisting of disabled and non-disabled members. Much of this work has emphasized the challenges that disabled team members face and the \textit{accessibility practices} used to support them, such as accommodations \update{including modifications to meet accessibility needs, the use of assistive technologies, as well as informal norms and behavioral adjustments adopted by others to provide support.} For example, Mack et al. \cite{mack_mixed_2021} identified how their virtual team struggled with remembering accessibility practices, navigating conflicting access needs, and negotiating the role of power dynamics in upholding norms. Alharbi et al. \cite{alharbi_accessibility_2023} demonstrated how hybrid meetings amplify these tensions through barriers such as difficulty recognizing speakers, degraded audio quality, and captioning errors. Although these studies provide valuable insights into accessibility barriers and practices, the broader impact of accessibility practices on how disabled and non-disabled team members work together remains underexplored. For example, it remains unclear how such practices shape collective understandings of productivity, whether accommodations generate interpersonal tensions, or conversely, foster stronger connections within the team, and how they influence the participation and engagement of all team members.

\textit{Teamwork}, the interactions and relationships that influence how a team functions and performs \cite{lee_conceptual_2004}, is central to understanding how effective collaboration is achieved in diverse team environments. Prior work in this field has studied teamwork across diverse groups, including multicultural teams \cite{cagiltay_working_2015}, socioeconomically diverse teams \cite{hofhuis_diversity_2016}, and linguistically diverse teams \cite{kalra_alike_2023}. These studies show how differences in background or communication styles affect coordination, participation, and rapport. For instance, members of multicultural virtual teams often experience friction due to divergent communication norms and expectations, which can simultaneously hinder cohesion and spark creativity \cite{cagiltay_working_2015, stahl_unraveling_2021}. These findings echo the broader “double-edged sword” characterization of team diversity: diversity introduces both positive and negative effects \cite{stahl_unraveling_2021}. Yet, despite growing recognition of these dynamics in cultural and other forms of diversity, mixed-ability teams remain largely overlooked in teamwork research. In other words, there is a gap in understanding how accessibility practices not only support disabled team members but also shape collective processes of productivity, participation, and camaraderie across the team.

To address this gap, we examined how accessibility practices influence teamwork in mixed-ability teams that collaborate virtually. Beyond focusing on individual experiences of access, we investigated how accessibility practices influence the broader functioning of teams. We aim to answer the following research question: 

\textbf{How do accessibility practices influence key aspects of teamwork, such as productivity, participation, and camaraderie, among disabled and non-disabled members of mixed-ability teams in virtual collaboration settings?} 

To answer this question, we conducted semi-structured interviews with 18 participants, including 12 with and 6 without disabilities, who work in mixed-ability teams that collaborate in virtual settings. Through these 60-minute interviews, we asked participants about their teams’ backgrounds and facets of teamwork, specifically productivity, participation, and camaraderie \cite{leshed_visualizing_2009, benk_is_2022, hackman_theory_2005, hackman1975group, woolley_evidence_2010}.

Our findings revealed that accessibility practices actively shaped how disabled and non-disabled team members coordinated tasks, sustained relationships, and negotiated responsibilities in mixed-ability teams that collaborate virtually. Accessibility practices shaped stronger team-wide processes, such as building rapport and maintaining productivity. However, accessibility practices also introduced challenges. Participants described emotional and interpersonal tensions, such as frustrations and challenges with balancing empathy, accommodations, and accountability. These tensions affected both disabled and non-disabled team members and hindered camaraderie among team members. Finally, non-disabled participants described allyship not as a fixed set of actions or formal training, but as a learning process and a teamwork skill that emerged through interaction with their disabled team members and was shaped by their team’s culture.

Our contributions are threefold. First, we demonstrate that accessibility practices are not merely a means of inclusion but mechanisms that actively shape rapport, productivity, and participation. Second, we uncover how accessibility practices shape the experiences of both disabled and non-disabled team members, surfacing underexplored emotional and interpersonal challenges. Third, we present \update{recommendations for team practices to address the challenges faced by participants and design opportunities for virtual collaboration tools to support those practices.} \update{These insights and recommendations for team practices can help mixed-ability teams improve their teamwork. They also highlight opportunities for HCI and CSCW researchers to design virtual collaboration tools that better support mixed-ability teams.} We reframe accessibility practices not just as a means of supporting people with disabilities but also as a pathway to fostering strong teamwork. 
\section{Related Work}
In this section, we bring together two bodies of research: teamwork and accessibility in mixed-ability virtual collaboration. We first review foundational work on teamwork, focusing on dimensions of teamwork effectiveness such as productivity, participation, and camaraderie, and how these dynamics unfold in diverse groups. We then examine research on accessibility in mixed-ability teams, emphasizing dynamics such as access labor, interdependence, stigma, and allyship in virtual collaboration. By linking these perspectives, we identify a gap in understanding how accessibility practices shape the core processes of teamwork in mixed-ability virtual collaboration settings.
\subsection{Teamwork} \label{dimensions}
Researchers have developed numerous frameworks that capture the multidimensional nature of teamwork \cite{leshed_visualizing_2009, benk_is_2022, hackman_theory_2005, hackman1975group, woolley_evidence_2010, salas_is_2005, tuckman_developmental_1965, hackman_leading_2002}. Across this extensive body of work, various dimensions consistently emerge as key to understanding effective teams, three of which are productivity, participation, and camaraderie \cite{kozlowski_enhancing_2006}. Kozlowski et al. \cite{kozlowski_enhancing_2006} reviewed over 50 years of research, revealing that these dimensions are fundamental to enhancing team effectiveness. Our work uses these dimensions as a lens to understand teamwork in mixed-ability teams. 

\textit{Productivity} reflects how effectively team members coordinate their efforts to achieve shared goals \cite{anders_team_2016, kozlowski_enhancing_2006}. Effective coordination not only ensures task completion but also reduces inefficiencies from misunderstandings or redundant efforts \cite{espinosa_team_2007, cheng_age_2023}. For instance, Scott et al. \cite{barker_scott_designing_2024} highlight that fostering positive relationships and a collaborative environment is essential for enabling team members to focus on common objectives.

\textit{Participation} is defined as the active engagement of team members in shared tasks and decision-making processes \cite{kozlowski_enhancing_2006}. Consistent participation is crucial for maintaining productivity and fostering cohesion. Research shows that balanced participation promotes engagement, supports shared responsibility, and prevents imbalances in workload or decision-making authority \cite{cohen_teamwork_1991, schittekatte_effects_1996, bowers_reflections_2001, woolley_evidence_2010}. For example, Woolley et al. \cite{woolley_evidence_2010} found that teams with uneven participation are less productive, while Malone et al. \cite{malone_interdisciplinary_1994} and Fischer et al. \cite{fischer_meta-design_2004} emphasized the need for environments where every team member can contribute meaningfully.

\textit{Camaraderie} involves building interpersonal connections and collective motivation \cite{kozlowski_enhancing_2006}. Valentine and Edmondson \cite{valentine_team_2015} noted that collaboration often creates opportunities for camaraderie alongside task completion. Similarly, Hackman and Morris \cite{hackman1975group} argued that camaraderie can boost collective motivation, enhancing overall team effectiveness. Degnen and Rosenthal \cite{tickle-degnen_nature_1990} described camaraderie as “mutual positivity,” a dynamic of friendly and caring behaviors that underpins effective teamwork \cite{nadler_rapport_2003}.

Prior work has examined how teamwork unfolds in diverse groups, including multicultural teams \cite{cagiltay_working_2015}, teams with varied socioeconomic backgrounds \cite{hofhuis_diversity_2016}, and linguistically diverse teams \cite{kalra_alike_2023}. These studies show that differences in background and communication styles shape collaboration in significant ways. For example, members of multicultural virtual teams often encountered challenges from divergent communication norms, which hindered coordination and task outcomes \cite{cagiltay_working_2015}, while Hofhuis et al. \cite{hofhuis_diversity_2016} demonstrated that cultural diversity could also promote trust and openness.

This literature reflects the broader characterization of diversity as a “double-edged sword”: it can foster creativity, innovation, and decision quality, yet it can also lead to conflict, miscommunication, and reduced cohesion \cite{stahl_unraveling_2021}. For example, meta-analyses confirm this duality, finding that culturally diverse teams tend to be more innovative but also more prone to conflict and lower cohesion \cite{horwitz_diversity_2007, stahl_unraveling_2021}. Despite this duality, frameworks in this field have often emphasized the challenges of diversity, such as in-group bias \cite{mannix_neale_2005}.

Our work builds on this body of research by extending the study of teamwork into the context of mixed-ability teams. While cultural, socioeconomic, and linguistic diversity have been studied, mixed-ability teams remain underexamined. The role of accessibility practices in shaping productivity, participation, and camaraderie has been underexplored, despite its potential to both introduce challenges and support teamwork.
\subsection{Accessibility in Mixed-Ability Virtual Collaboration}
Accessibility research highlights several dynamics that shape collaboration between disabled and non-disabled team members, including access labor, interdependence, stigma, and allyship. \textit{Access labor} refers to the additional work disabled team members undertake to navigate inaccessible systems, often falling disproportionately on them \cite{branham_invisible_2015, das_towards_2021, borsotti_neurodiversity_2024}. Such work can be both visible and invisible, ranging from educating team members about accessibility to developing workarounds \cite{branham_invisible_2015, das_towards_2021}. Scholars stress that technology alone cannot mitigate these burdens; inclusive social practices are equally essential \cite{mack_mixed_2021, alharbi_accessibility_2023}. Studies also highlight the role of \textit{allyship}, where non-disabled team members advocate for their disabled team members to make environments accessible, though disabled team members continue to shoulder most of this responsibility \cite{mack_mixed_2021, alharbi_accessibility_2023}.

Mixed-ability collaboration also depends on \textit{interdependence}, the ways members rely on each other to coordinate support and achieve collective goals. Prior work frames interdependence not as dependency but as an essential resource for sustaining equitable participation \cite{kameswaran_we_2018, soro_beyond_2019, yildiz_virtual_2023}. At the same time, team members sometimes conceal disabilities to avoid \textit{stigma}, the negative stereotypes that threaten their acceptance in a group, particularly among those with chronic illness or neurodivergence \cite{shinohara_shadow_2011, branham_invisible_2015, ganesh_work_2021, lowy_toward_2023}. 

Prior work has explored these dynamics in virtual collaboration contexts. Through an autoethnography of their virtual team, Mack et al. \cite{mack_mixed_2021} demonstrated that while remote collaboration created opportunities for participation, it also surfaced tensions when different accommodations conflicted. Similarly, Yıldız and Subaşı \cite{yildiz_virtual_2023} examined a mixed-ability team in Turkey and highlighted how virtual platforms both supported interdependence and exposed inequities in access, underscoring the need for collective responsibility of access in virtual workspaces. Alharbi et al. \cite{alharbi_accessibility_2023} studied accessibility in hybrid meetings with professionals with disabilities, revealing not only barriers, such as poor audio quality and conflicting access needs, but also repairs, where access labor was shared among team members to negotiate tensions and co-create access. \update{Kaschnig et al. \cite{kaschnig2024remote} explored remote collaboration between sighted and non-sighted participants who worked together in teams to complete problem-solving tasks. Their findings show that non-sighted participants encountered accessibility issues even when using screen readers, and that team members addressed these challenges through shared support. Sighted participants described visual information to their non-sighted team members, while non-sighted participants assisted each other to work around accessibility barriers.} Together, these works suggest that ensuring accessibility in mixed-ability virtual teams requires not only technological solutions but also social and cultural practices.

Prior work examining specific disability groups further illustrates these challenges in virtual meetings. Leporini et al. \cite{leporini_distance_2021} evaluated video conferencing platforms for blind participants, documenting persistent screen reader and keyboard accessibility gaps that hindered participation. Cha et al. \cite{cha_you_2024} explored the accessibility of virtual meetings for blind and low vision (BLV) professionals, showing how inaccessible meeting tools required extra labor, discouraged disclosure of disability, and often forced participants to develop ad-hoc fixes. Akter et al. \cite{akter_if_2023} further examined BLV meeting facilitators and found that inaccessibility in hosting features constrained leadership opportunities and required extensive preparation and coordination. \update{Kim \cite{kim_recognizing_2024} found that sighted participants often misinterpreted facial expressions of BLV people in Zoom meetings, highlighting how the virtual setting can distort emotional communication.} These studies suggest that virtual settings shape accessibility, social participation, disclosure practices, and perceptions of professional competence.

Prior work also highlights how virtual collaboration impacts neurodivergent and autistic team members. Zolyomi et al. \cite{zolyomi_managing_2019} found that autistic adults often experienced heightened stress during virtual meetings due to sensory sensitivities and cognitive load, requiring them to adopt coping strategies such as masking or minimizing sensory input. Das et al. \cite{das_towards_2021} examined neurodivergent professionals’ experiences working from home during the pandemic, finding that they had to continuously negotiate between productivity and wellbeing while creating accessible physical and digital workspaces. These findings complement Tang’s interviews \cite{tang_understanding_2021} with disabled teleworkers, which revealed how design choices in video calling and collaboration tools (e.g., active speaker detection, screen sharing) inadvertently excluded certain groups or exposed disabilities. Collectively, this body of work suggests that virtual collaboration tools must move beyond baseline accessibility features toward supporting diverse needs.

The prior work reviewed here suggests that accessibility practices not only shape individual participation but also influence the collaborative dynamics and coordination of teams. However, most existing research has focused on individual experiences or specific tools, with less exploration of how mixed-ability teams collectively navigate the core aspects of teamwork. Building on this foundation, our work provides an \update{account} on how accessibility practices impact teamwork in mixed-ability teams that collaborate virtually.
\section{Method}
To address our research question, we conducted semi-structured interviews to capture the nuanced experiences of participants in mixed-ability teams and to gain qualitative insights into their teamwork practices, experiences with accommodations, and use of virtual meetings and tools. This approach allowed us to explore the general processes underlying teamwork in virtual settings. 
\subsection{Participant Recruitment and Demographics}
\input{participants}

To understand teamwork from diverse perspectives, we recruited disabled and non-disabled participants who work in mixed-ability teams. We required participants to be part of teams that regularly hold virtual meetings (i.e., had enough experience to talk about virtual meetings). We sought participants from teams of any type or size across any field. Participants’ collaboration in virtual settings extended beyond these synchronous meetings to include broader virtual practices through asynchronous tools. We recruited participants through mailing lists, social media sites, and snowball sampling. This study and recruitment materials were approved by our university’s Institutional Review Board.

We screened potential participants’ eligibility through a survey. In our screening process, participants self-identified their disabilities, and we ensured that any disclosure was voluntary. Additionally, to respect privacy and considerations around disability status, we clarified that participants were not required to disclose the specific disabilities of their team members. Instead, participants were only asked to confirm whether their teams included members with disabilities. This approach allowed us to identify mixed-ability teams without requiring personal or detailed information about other team members’ disabilities.

We recruited 18 participants (4 male, 14 female). 12 participants identified as having a disability (e.g., BLV, physical disabilities, \update{attention-deficit/hyperactivity disorder (ADHD)}, Autism, chronic illness). 6 participants did not identify as having a disability. Participant ages ranged from 19 to 56 years (mean = 32.55). Participants were a part of \update{different} mixed-ability teams with people with (e.g., BLV, \update{deaf and hard-of-hearing (DHH)}, physical disabilities, ADHD, Autism, chronic illness) and without disabilities. Participants used a variety of virtual meeting platforms, such as Zoom, Google Meet, and Microsoft Teams. Participants also used various tools to collaborate with their teams, such as Discord, Slack, Miro, Google Calendar, Google Drive, Emails, and Figma.

Participants were engaged in various professional and academic roles across fields such as education, nonprofit work, technology, and research. Teamwork settings varied; some participants worked in small teams of a few members, others in medium-sized teams of around 8-20 members, and a few occasionally participated in larger teams with up to 70 people. These mixed-ability teams engaged in diverse collaborative tasks relevant to their fields, including research, corporate work, support services, and instructional work. 

Table \ref{tab:participant-demographics} shows detailed participant demographics. We refer to participants without disabilities as N\# and participants with disabilities as D\#. 
\subsection{Procedure}
We conducted a semi-structured interview with each participant that lasted approximately 60 minutes, held via Zoom and audio-recorded. Participants were compensated with a \$30 gift card for their time. 

We began the interview by asking demographic questions and participants' backgrounds regarding their teams. We asked about their teams’ use of virtual meetings and other tools to support collaboration in virtual settings. We also asked the participants to describe the members of their team (e.g., disability status and role on the team). We then asked about accommodations, such as the use of assistive technologies, in participants’ teams. Specifically, we asked about how they were formalized, practiced, and maintained. 

To explore teamwork in mixed-ability teams, we drew on established literature to identify key dimensions of teamwork. Given the diversity and complexity of existing frameworks, choosing a focused yet comprehensive approach was a significant challenge, as noted in prior work \cite{samrose_coco_2018, kozlowski_enhancing_2006}. No single framework fully encapsulated the intricacies of teamwork in virtual, mixed-ability contexts. Thus, we focused on three interrelated dimensions frequently identified in the literature—productivity, participation, and camaraderie—based on a review by Kozlowski et al. \cite{kozlowski_enhancing_2006}.

We developed sets of questions to ask about these dimensions of teamwork, and offer some examples below:
\begin{itemize}
\item{Productivity (e.g., How do you feel about your team’s productivity?, 
\update{How do you collaborate with your team members to complete tasks?})}
\item{Participation (e.g., What’s your perception of the level of participation among yourself and your team members?, \update{How comfortable are you with expressing your opinions during meetings?})}
\item{Camaraderie (e.g., Have you ever felt connected to [team member]?, \update{Can you give an example of a time when you felt particularly connected to [team member]?})}
\end{itemize}

We asked about participants' perceptions of themselves and their team members for each dimension. We carefully worded our questions to encourage honest and thoughtful responses without feeling pressured to conform to socially acceptable narratives. \update{Throughout the interview, we also elicited concrete details around specific instances and examples. For example, we asked open-ended follow-up questions such as, “Can you describe a particular experience working with [team member]?” and “Can you recall a recent instance when…,” which encouraged participants to describe concrete experiences. We also asked follow-up questions to probe details about these experiences.} We were intentional in using inclusive language that does not attribute a team member’s disability as a deficit.

Finally, we asked participants to reflect on teamwork in mixed-ability teams (e.g., What do you think makes for good teamwork with your team?, How do disability, accessibility, and accommodations affect your mixed-ability team?). \update{For reference, the semi-structured interview protocol can be found in Appendix A.}
\subsection{Data Analysis}
\update{We audio-recorded and transcribed all interviews. While our interview protocol structure was informed by prior work on key dimensions of teamwork, these dimensions were not used to create predefined codes for deductive analysis. The coding process was inductive and data-driven, in which we applied open descriptive codes to participants’ experiences rather than imposing predefined categories onto the data \cite{clarke_thematic_2017, braun_thematic_2021}. While our coding process was inductive, we align with Braun and Clarke's view that some elements of deductive reasoning may have been interwoven with our inductive approach \cite{braun2006using}.} 

\update{Three researchers each coded two transcripts separately, then came together to discuss the codes. Through collaboratively discussing codes, we settled on a preliminary codebook, which included descriptive open codes reflecting participants’ accounts and experiences (e.g., accessibility as shared team responsibility, importance of feeling safe and comfortable discussing needs). We then split up the remaining transcripts, continuing to meet regularly to discuss newly emerging codes and consolidate overlapping ones.}

\update{For thematic analysis, we clustered related codes on a shared digital board to create affinity diagrams and identify themes and subthemes across codes. Throughout data collection and analysis, the research team engaged in iterative discussions, which led to consensus on the final set of themes. Examples of quotes and applied codes can be found in Appendix B.} 
\section{Findings}
\subsection{Accessibility Practices Enhance Rapport, Coordination, and Productivity}
\subsubsection{Sharing and Establishing Accessibility Needs Together as a Team Builds Rapport and Reduces Stigma}
Participants described how team-wide engagement with accessibility as a shared and ongoing practice built rapport within their teams and reduced stigma. Participants described how their teams did not treat accessibility as an individual responsibility but rather as a shared responsibility of the whole team. Their teams regularly discussed and treated accessibility as a collective concern, which helped normalize conversations about disability and fostered a \textit{“safe”} and \textit{“comfortable”} environment (D9). Being able to express their needs in these environments allowed participants to build stronger interpersonal relationships with their team members.

For example, D10 described how consistent check-ins about accessibility fostered rapport, contributing to a team culture that felt like family: \textit{“The way we meet is like a family; we check in and always check on each other to find out if people are okay and have the accommodations they need.”}

Some participants described formal mechanisms in their teams to establish accessibility practices \update{using virtual tools such as shared online documents and online surveys to support transparency and sustain accessibility conversations}. D6 described a practice in her team where both disabled and non-disabled team members documented accessibility needs and personal needs in a shared online document. This practice built rapport, reduced stigma, and helped team members align expectations and support one another:

\textit{\begin{quote}
“We usually have a sheet that everybody writes specifically what accommodations they may need or anything else in their life that may contribute to their teamwork on the team. I think that just helps build rapport within the team as well as destigmatize. And just gives everybody kind of like a mental model for what to expect from each person during the meetings.”
\end{quote}}

Similarly, D11 described how her team used a shared survey to communicate needs ranging from disability-related needs to family and caregiving responsibilities. For instance, she used this survey to disclose that \textit{“one of [her] family members was going through chemotherapy”} and that she would need time off for caretaking. Others, she noted, used it to indicate scheduling needs related to childcare or mental health. By making these needs visible, the survey helped the team better understand and support both disabled and non-disabled members.

In teams where conversations about accessibility were not yet a cultural norm, participants appreciated when others took the initiative to raise conversations about accessibility. D5 explained that when team members initiated accessibility practices, it reduced stigma and created a more inclusive culture: \textit{“I don’t think I would think to ask, ‘Hey, can we do this?’ But someone else initiating it would normalize it. Like, oh, that person asked. It’s a thing, a norm, a cultural shift.”}
\subsubsection{Developing Shared Understanding of Access Needs Supports Coordination and Productivity}
To stay productive, participants’ teams relied on both individual reflection and a shared understanding of how disability affected each person’s work style, pace, and needs. \update{Since participants collaborated virtually, understanding one another’s access needs was especially important for coordinating work across different schedules, tools, and platforms.} This shared understanding, built through open communication, allowed teams to coordinate more effectively, align expectations, and support one another’s productivity.

For some participants, developing this shared understanding began with self-reflection. D6, for instance, described how recognizing her own disability and identifying the right accommodations helped her contribute more effectively to the team:

\textit{\begin{quote}
“Disabilities definitely have an impact on productivity. But I think knowing how your disability manifests helps with guiding the accommodations to make you more productive. The person (team member) needs to know how their disabilities affect them and their productivity, and the accommodations they need… the accommodations should help with productivity.”
\end{quote}}

By understanding how her disability, ADHD and Autism, affected her productivity, D6 was able to identify what support she needed. \update{The virtual collaboration setting allowed D6 to try different accommodations and determine which ones supported her need for communication clarity. For example, she found that using live transcripts, rewatching meeting recordings, and participating through the chat were especially helpful.} This individual awareness served as a basis for communicating those needs to the team and participating more fully.

While D6 emphasized self-awareness, N3 spoke to the importance of understanding and being attentive to her team members’ productivity needs. She highlighted the importance of knowing \textit{“how each person can be productive and can be effective to the team.”} She explained that as a non-disabled team member, it was important to ensure that the needs of her team members are \textit{“effectively communicated to the rest of the team.”} N3 emphasized that this kind of shared understanding not only helped individual team members work effectively but also made it easier for her to be a supportive team member.

For some participants, these conversations about productivity needs resulted in team members learning about and adopting accessibility practices from each other. D9, who has ADHD, described how she and another disabled team member with ADHD began using the concept of “crip time” to manage deadlines in their team. \update{The term “crip time,” which originated from disability studies, was explained by Kafer \cite{kafer2013feminist} as flexible, non-linear experiences of time that challenge normative expectations of productivity and pace. Rather than viewing time as fixed or uniform, crip time recognizes that disabled people may need to work at different timelines in response to fluctuating access needs \cite{kafer2013feminist}.} D9 initially used the term crip time to describe her flexible approach to timelines, acknowledging that her pace of work might vary due to her disability. Her team member then adopted the same language and approach, and together they began using crip time as a shared framework for negotiating flexible deadlines. This practice shaped how D9 and her team member communicated about productivity and expectations.
\subsection{Accessibility Practices Can Result in Emotional and Interpersonal Tensions}
\subsubsection{Compensating for Inaccessible Tools Strains Camaraderie and Creates Emotional Burden for Non-Disabled Team Members} \label{inaccessible-tools}
Participants described how the use of inaccessible tools created challenges that affected the entire team, not just disabled members. Non-disabled participants frequently took on additional work to compensate for these barriers, leading to unequal distributions of labor and emotional strain around roles, fairness, and responsibility. \update{For example, D4, who is blind, explained that her team’s database was not screen-reader accessible, which resulted in sighted team members completing data-entry tasks that blind team members could have completed \textit{“if we had the right software.”}}

Several non-disabled participants reflected on the toll this dynamic took on their emotional well-being and on their relationships with team members. N1 described her willingness to step in regularly to sustain team productivity when progress was slowed by accessibility barriers: \textit{“If things need to speed up, [we] can be of service and be a catalyst.”} While N1 was eager to be supportive, she noted that her role on the team increasingly involved taking on additional responsibilities.

Other participants, such as N8 and N12, expressed discomfort with the consequences of taking on more work. N12, for example, shared complex feelings about how inaccessible tools shaped his role on the team and strained relationships with his team members. \update{N12, who works with BLV team members, explained that their team’s primary system, Salesforce, is \textit{“not the most accessible… when you’re using [it] non-visually.”} He contrasted his own speed with the slower, multi-step navigation required with a screen reader: \textit{“It takes my coworkers longer… they have to go through so many menus… and you have to wait for the screen reader to read the label… before you can move to the next one.”} As a result, he described how he completed tasks faster than his BLV team members: \textit{“I can do stuff faster on Salesforce… which means I could get things done faster, and then I could do extra admin stuff on top of that.”}}

N12 desired to be helpful, but felt burdened by the expectation that he should take on more work because he could complete tasks more efficiently: \textit{“If you do it too fast, you don’t want everyone to know... I feel like more stuff does get dumped on me because of it.”} N12 demonstrated an internal tension between the desire to be supportive and a growing sense of frustration over inequity. This tension contributed to emotional fatigue and discomfort when interacting with his team members. 

Interestingly, N12 emphasized how his BLV team member proactively checked in to make sure he was not overwhelmed. This gesture of support helped alleviate some of the pressure N12 was feeling and demonstrated a reversal of the dynamic often expected between disabled and non-disabled team members, where support typically flows from the non-disabled to the disabled team member.

\subsubsection{Accommodating Team Members’ Needs Raises Tensions Between Empathy and Accountability} \label{accountability}
Participants faced significant tensions in balancing empathy, accommodations, and shared accountability with their team members. Although flexibility and empathy were viewed as essential for supporting accessibility needs, several participants expressed frustration that it often resulted in delays and uneven contributions. These challenges were not limited to individual team members but had broader consequences for the team’s productivity and morale.

For example, D6 described the difficulty of navigating the fine line between \textit{“being cautious and empathetic”} and \textit{“holding people up to their responsibilities.”} She expressed frustration when team members used accommodations but did not follow through on their tasks. In one instance, a team member consistently failed to complete work on time. D6 recounted how she and other team members had to meet separately to discuss how to move forward in addressing the issue. They ultimately decided to give the team member another chance, but D6 found the situation deeply frustrating. It felt, to her, as though the team member was using their disability as an excuse to avoid accountability.

Participants shared how they made consistent efforts to be empathetic and accommodating toward their team members. However, these efforts often became emotionally taxing and impacted their own productivity. Over time, participants grew frustrated and overwhelmed by the ongoing burden of compensating for others’ lack of accountability, even when accommodations were provided. \update{In virtual collaboration, these tensions were compounded by the reliance on remote reminders and coordination, which could be overlooked or taken lightly.} N8, for example, noted the recurring need to send reminders to ensure team members completed their tasks or attended meetings. D11, who often collaborated with team members with ADHD, echoed this frustration: \textit{“Honestly, I don't think I should have to remind someone to show up to a (virtual) meeting.”}

Despite their efforts, participants did not report successful strategies for resolving these tensions. In D11’s case, the ongoing frustrations around accountability ultimately led her to stop working with certain disabled team members. She explained that collaborating with them had become too stressful, and she preferred to work with other team members instead.
\subsubsection{Disabled Team Members Face Emotional Strain From Managing How They Use Accommodations to Avoid Judgment} \label{accountability-disabled}
Disabled participants described internal tensions around how and when to use accommodations while striving to be seen as accountable, respectful, and reliable members of their teams. Rather than using accommodations solely as accessibility supports, participants also viewed them as social tools they had to manage carefully to avoid being misjudged or perceived as irresponsible by others. This meant that accommodations were not just about enabling participation, but also about managing how they were perceived within their teams.

Participants wanted their teams to see them as accountable, even while they relied on accommodations to manage disability-related challenges when collaborating virtually. For some, this meant navigating internal tensions about when and how to use accommodations without feeling like they were making excuses and being seen as irresponsible. D11 described a constant effort to balance accommodations with a sense of responsibility:

\textit{\begin{quote}
“There's this fine line of being flexible and accommodating with your ADHD, and then using it as a form of a lack of accountability. I am very flexible with myself, but I also don't want to be someone who makes excuses a lot.”
\end{quote}}

This kind of internal monitoring placed an emotional burden on participants, especially in teams where accessibility was not already normalized. In these environments, participants worried that requesting accommodations might negatively affect how team members judged their professionalism or reliability. D9, for instance, described feeling more at ease disclosing access needs when working with other disabled team members, who she trusted not to interpret her requests as laziness or irresponsibility:

\textit{\begin{quote}
“It's often with another team member who is also disabled, or with a team that's predominantly disabled, in which I feel comfortable… to speak up about these access needs, to raise access concerns as they come across, and not have them think of these things as me being a lazy or a bad [team member].”
\end{quote}}

By contrast, in teams where accessibility was not a shared norm, D9 felt hesitant to advocate for her needs out of fear of being judged: \textit{“It’s definitely easier when we already have an established norm around… asking for access needs. In other spaces, it feels more judgmental… It’s not very safe to bring up an access concern.”} Participants modulated their use of accommodations not only to manage their own productivity, but also to protect how they were perceived by others. This constant self-regulation, driven by concerns about being seen as an equal and reliable team member, took an emotional toll on disabled team members. 
\subsection{Accessibility Practices Support Allyship and Relationships Between Team Members}
\subsubsection{Non-Disabled Team Members Developed Allyship Through an Ongoing Learning Process} \label{allyship-learning}
Non-disabled participants described allyship, the practice of supporting and advocating for their disabled team members, as a skill they learned over time through being part of inclusive teams. Rather than relying on formal accessibility training, participants developed their approach to allyship through observation, ongoing reflection, and responsiveness to informal cues within the team environment. Participants described thinking critically about subtle cues they may miss when collaborating virtually, how their actions affected others, and adjusting their behaviors to better support their disabled team members. Supportive team cultures that emphasized openness, psychological safety, and shared responsibility for access allowed participants to engage in this process of reflection and adaptation, helping them become more attentive and responsive team members. Importantly, participants framed allyship not only as a form of social support but also as a teamwork skill that directly enhanced coordination and productivity by enabling smoother collaboration.

For N1, allyship was not about following a predefined set of actions, but about shared learning. She explained, \textit{“Everybody is a learner and also a teacher in some form,”} describing allyship as a shared, evolving process in which team members learn from one another through working together. This framing shaped how she approached her role on the team: rather than relying on fixed assumptions, she emphasized the importance of paying attention, adjusting her behavior, and remaining open to feedback. Through this process, she learned how to be a more supportive ally over time. One shift she described was learning to step back in virtual meetings to create space for others to contribute. She noted, \textit{“It's important to take the time to make that space for [team members with disabilities] to talk,”} especially during virtual interactions. \update{N1 recognized that in virtual meetings, her disabled team members with physical disabilities may face additional barriers when trying to contribute due to difficulties signaling when they want to speak or delays in unmuting or navigating tools.} For her, stepping back was not only about being respectful but also about improving the flow of team discussions, ensuring that contributions were more evenly distributed and coordination of ideas happened more smoothly.

For N16, learning to be an ally involved navigating unfamiliar norms, which required cognitive and emotional effort. She joined a team in which American Sign Language was the primary mode of communication, a setting that was unfamiliar to her. Without prior experience in DHH cultural contexts, she found herself navigating unspoken expectations that she had not previously encountered. Upon learning that some of her early behaviors may have been inappropriate, she expressed concern about unintentionally causing offense. As she explained, \textit{“even though I'm trying not to be disrespectful, there are some things I had to learn with time and with experience being around the DHH community.”}

N16 did not receive formal training on how to best collaborate with her team members. At first, she found it challenging to participate because many of the norms, such as signing instead of clapping or visually signaling agreement, were unfamiliar and not explicitly introduced. She described the early stages as confusing and sometimes uncomfortable, since she worried about unintentionally making mistakes. Over time, however, she adapted by observing others, asking questions, and modifying her behavior. \textit{“Even though we're not fluent signers... It's just something you pick up on,”} she explained. She emphasized that understanding and adopting these practices required ongoing attention and effort, especially when interacting with her team members virtually. By gradually learning and incorporating these signals, she was able to overcome initial difficulties, reduce the chance of miscommunication and disruption, and help the team sustain clear communication and remain productive.

N16 described how her team’s inclusive environment played a critical role in supporting this learning process. Although it was uncomfortable for her at first, her team was open to questions and responsive to mistakes, which allowed her to continue adapting her behavior. She stated, \textit{“I would say this is a safe learning space... I just had to remind myself that I’m here to learn.”}
\subsubsection{Non-Disabled Team Members Faced Challenges in Determining When and How to Offer Help without Overstepping} \label{allyship-overstep}
Participants described allyship as an ongoing process of figuring out how to offer support without overstepping. Non-disabled team members, in particular, reflected on the challenges of supporting disabled team members in ways that respected their autonomy. When collaborating virtually, participants were more cautious about offering help, as it often required overt actions, such as sending a message or interrupting verbally, which could feel more disruptive than in person. Rather than assuming that offering help was inherently positive, participants described learning to treat allyship as something that needed to be negotiated both within themselves and in conversation with others.

For some, this involved learning to manage the instinct to step in. For example, N12 explained that he initially viewed allyship as actively looking for ways to help, but came to recognize that this could feel intrusive or ableist. He reflected:

\textit{\begin{quote}
“I was always looking for ways to help, and I realized that was a form of ableism... It's important to let people be themselves and do what they need to do without assuming anything about them or anything that they need.”
\end{quote}}

This realization emerged through a conversation with a disabled team member who shared that they sometimes preferred to work through tasks independently, even if it took longer. N12 recalled: \textit{“I asked them, ‘Hey, does it bother you that I'm always kind of just stepping in and helping?’ And they said, ‘Yeah, you know, sometimes I just would like a little time... I'm figuring it out.’”}

This exchange stayed with him and influenced how he approached interactions going forward. Rather than intervening immediately, he described learning to pause and give space: \textit{“If I get the urge to jump in, I stop myself and just say, ‘Okay, let’s just see...’ And then if they get stuck or if they’re not progressing, I just... start talking out loud... ‘What are our options?’”} For N12, allyship required restraint, patience, and an ongoing process of learning when his support would be helpful.

Another participant, N18, reflected on the emotional and interpersonal complexity of being an ally. As a sighted person working closely with BLV team members, she described her approach to allyship as grounded in empathy but also shaped by a clear awareness of her limitations. She explained, \textit{“as a sighted person I can have... empathy, but I don't have that personal knowledge of what it's like to live with the disability.”} This recognition informed how she remained cautious not to impose her assumptions when offering support.

She emphasized that allyship required careful attention to how help might be received. Offering assistance could sometimes be perceived as patronizing or overstepping, even if the intention was well-intended. As she put it, \textit{“you don't want anybody to feel like you're trying to take care of them, or like they don't know how to do things for themselves... that kind of overstepping we definitely want to avoid.”} At the same time, she acknowledged that support can be both needed and appreciated when it is offered in a way that respects autonomy. To navigate this tension, she described adopting a strategy of asking first: \textit{“Usually, the best thing is to just ask, ‘Do you want assistance with that or not?’ rather than assuming.”}

N18 also described how working alongside disabled team members with lived experience helped her better understand accessibility-related challenges she might otherwise overlook. She noted that conversations with team members often surfaced details she had become desensitized to, explaining that \textit{“there's just things that you don't think about... and so they'll bring up a point... that'll be something that I hadn't thought of.”}
\section{Discussion}
Our findings demonstrate that accessibility practices shape the emotional, social, and practical dynamics of teamwork, including productivity, participation, and camaraderie, for both disabled and non-disabled team members. \update{Although we asked participants about their virtual collaboration experiences, some of the experiences and challenges they described were not specific to the virtual collaboration setting. Instead, they reflected broader dynamics of teamwork, such as navigating the learning process of allyship or building rapport.} By centering the underexplored dynamics of collaboration between disabled and non-disabled people, our work highlights that beyond providing access for disabled team members, accessibility practices shape team-wide processes such as rapport-building, productivity, participation, and perceptions of fairness and allyship.

\update{The challenges uncovered in this work must first be addressed through human practices, such as by setting team norms and building supportive team cultures. At the same time, HCI and CSCW research have long shown that the design of tools can influence and support how people behave and interact. For example, participants already used existing virtual tools, such as shared documents and surveys, to communicate their accessibility needs. Similarly, there are opportunities for virtual collaboration tools to be designed to support these team practices, such as by prompting teams to check in with each other.}

\update{Below, we present the teamwork challenges uncovered in our findings, recommendations for team practices to address those challenges, and design opportunities for virtual collaboration tools to support these team practices.} We also discuss the limitations of our study and outline directions for future research.
\subsection{Addressing Teamwork Challenges: Recommendations for Practices and Opportunities for Tools to Support Them}
\input{design-rec} \update{Below, we outline the challenges uncovered in our findings and provide recommendations for team practices to address those challenges. These practices can provide directions for members in mixed-ability teams for stronger teamwork. We also outline design opportunities for virtual collaboration tools to assist these practices, which can inform HCI and CSCW researchers and developers in creating tools that support mixed-ability collaboration. These insights are summarized in Table \ref{tab:design-rec}.} 
\subsubsection{Uncovering the Burdens of Non-Disabled Team Members}
\update{\textbf{Challenge.}} While the impact of inaccessible tools on disabled team members is clearly established in prior work \cite{mack_mixed_2021, alharbi_accessibility_2023}, our findings show that unresolved inaccessibility also creates emotional and relational challenges for non-disabled team members (see Section \ref{inaccessible-tools}). Participants described how inaccessible tools and systems led to uneven task distribution, with non-disabled team members taking on extra responsibilities to sustain team progress. Although this labor was often framed as allyship, it introduced feelings of frustration, perceptions of inequity, and strain on interpersonal relationships, especially when the burden persisted without structural support or acknowledgment.

Prior work has documented the phenomenon of shared access labor in mixed-ability teams \cite{shinohara2011shadow, branham_invisible_2015, das_towards_2021, borsotti_neurodiversity_2024} and organizational literature has noted that invisible labor and unbalanced workloads can erode trust and engagement \cite{hackman_leading_2002}. Our findings extend this research by showing that inaccessibility also shapes the emotional dynamics of teamwork. Participants described how the repeated emotional labor of compensating for inaccessible systems left them feeling depleted, leading to distancing from their team members.

Importantly, we found that when teams openly acknowledged these relational impacts, they could foster reciprocal care. Disabled team members offered emotional support to their non-disabled team members, validating their frustrations or finding ways to reduce the load. These moments reconfigured allyship from a one-way act of support to a shared relational practice. 

\update{\textbf{Recommendation for Team Practice.} Teams can reduce uneven workload and emotional strain by openly discussing how accessibility barriers affect both disabled and non-disabled team members’ workloads. When accessibility issues arise, the team can decide together how to redistribute tasks instead of leaving non-disabled members to complete extra work without negotiation. Teams can openly review what slowed progress and who took on extra work together. Making these hidden efforts visible helps prevent frustration from building over time. It also reinforces the idea that accessibility is a shared team responsibility, not the burden of non-disabled team members taking on extra work when accessibility issues arise.}

\update{\textbf{Design Opportunity.}} 
\update{Virtual collaboration tools can help teams carry out these practices by helping teams surface hidden work and prompt shared task negotiation when accessibility barriers arise.} 

\update{Tools could include flags that let team members note access barriers (e.g., “Tool X caused a delay,” “Screen reader could not parse document”). These flags can create shared visibility about accessibility barriers, so task redistribution becomes a team conversation.}

\update{Tools can help team members highlight when tasks are reassigned due to accessibility issues. Existing platforms, such as Trello \cite{trello}, Asana \cite{asana}, and Jira \cite{jira}, allow users to assign and track tasks. Building on these platforms, tools could highlight when tasks are reassigned (e.g., “Reassigned to [team member] due to screen reader issue”). Tracking these reassignments can help teams review who took on extra work and why.}

\update{Tools can also prompt teams to engage in reflections about impact on workloads. For example, a weekly prompt could ask, “Did anyone take on unexpected extra work due to access barriers this week?” or “Do we need to revisit how tasks are distributed?” This would encourage teams to regularly reflect on workload balance.}

\update{Emotional awareness tools, such as mood check-ins or reflective prompts, similar to the Slack plug-ins such as Moodbit \cite{moodbit-slack}, could help teams surface emotional strain. These tools could create opportunities for shared conversations about team members’ frustrations.}
\subsubsection{Co-Creating Norms for Balancing Flexibility and Accountability}
\update{\textbf{Challenge.}} Our findings highlight a tension in mixed-ability teams: while accommodations are often offered with empathy, they can unintentionally create frustration and emotional strain when not accompanied by clear expectations for accountability (see Section \ref{accountability}). Participants described instances where they offered flexibility and accommodations to their team members, such as deadline extensions, but the recipients were not always accountable for follow-through. Without explicit norms, participants were unsure how long to extend flexibility, grew frustrated when contributions were delayed, and felt burdened by the additional work of checking in. Rather than strengthening inclusion, accessibility practices became a site of tension when expectations were not established.

Prior accessibility work has emphasized the value of flexibility in inclusive teams \cite{branham_invisible_2015, das_towards_2021, borsotti_neurodiversity_2024, mack_mixed_2021, alharbi_accessibility_2023, kameswaran_we_2018, soro_beyond_2019, yildiz_virtual_2023}, but these strategies often rely on informal norms. In contrast, research on teamwork and distributed collaboration highlights the need for role clarity, mutual accountability, and shared mental models to support effective coordination \cite{kozlowski_enhancing_2006, hackman_leading_2002, salas_is_2005}. Our findings surface a gap between these two bodies of work: flexibility enables access, but it is not sustainable without structures for shared accountability. Importantly, we found that participants valued flexibility and empathy, but they must be accompanied by deliberate team practices that define what accommodations mean, how responsibilities shift, and when to revisit expectations. In short, empathetic teamwork practices must be accompanied by a shared operationalization of accountability. 

\update{\textbf{Recommendation for Team Practice.} Teams should co-create explicit norms that ensure accountability remains in place even when accommodations are provided. Teams need shared expectations around how access-related needs will be supported, tracked, and followed through. Research on shared mental models \cite{salas_is_2005} highlights that clarity in roles and expectations improves performance. In practice, this can mean defining how long flexibility will last, what follow-up is expected, and when to check in again rather than relying on open-ended extensions or informal exceptions. For example, team members might agree that a deadline extension will be revisited at the next weekly meeting. Documenting these agreements, whether in shared documents or meeting notes, may help everyone stay aligned and accountable. Disabled participants also described pressures to demonstrate accountability despite requesting accommodations (see Section \ref{accountability-disabled}). These explicit norms also help disabled team members demonstrate their accountability and show that they are meeting expectations around flexible deadlines. By normalizing these conversations and negotiating flexibility and accountability together, teams can maintain both empathy and reliability, reducing uncertainty for all team members.}

\update{\textbf{Design Opportunity.}} 
\update{Virtual collaboration tools can support the team practice of establishing and revisiting explicit norms around flexibility and accountability.}

\update{Current collaboration platforms already provide some scaffolding for accountability. For example, Microsoft Teams Copilot \cite{microsoft_use_copilot_in_teams_meetings_2025} produces intelligent recaps that summarize key discussion points, suggest action items, and assign follow-ups, and Zoom’s AI Companion \cite{zoom_ai_assistant_2025} generates meeting summaries and highlights next steps for participants. These features help teams remember decisions and monitor tasks, which helps team members be responsible for their commitments.} 

\update{Building on this foundation, tools could make agreements around flexibility visible. For example, tools could allow team members to label a task with a “flexibility offered” tag and specify what follow-up is expected. This supports the team practice of co-creating explicit norms by documenting what was negotiated rather than relying on memory or informal agreements. Tools could also include temporal markers or  “flexibility timers” that indicate when a deadline extension is active and when the team agreed to check in again (e.g., “Revisit on Wednesday”). These markers can help team members track what types of flexibility were offered and agreed upon.}

\update{Additionally, tools could prompt teams to revisit expectations when a flexible deadline has been active for longer than planned (e.g., “This task’s flexible deadline has passed. Would you like to follow up or renegotiate?”). This supports the team practice of maintaining shared expectations.} 
\subsubsection{Encouraging the Learning Process of Allyship through an Inclusive Team Culture}
\update{\textbf{Challenge.}} Prior accessibility research has emphasized the importance of access as an inherently relational practice \cite{bennett_interdependence_2018}. Building on this foundation, our findings demonstrate that allyship is not only relational but also emerges as a teamwork skill that evolves through sustained interaction, reflection, and adaptation within the team context (see Section \ref{allyship-learning}). Prior work on allyship, largely from psychology, education, and social justice, has framed it as acknowledging systemic bias and privilege, grounding action in personal values, engaging in sustained high-effort practices, and supporting marginalized groups \cite{pietri2024framework, de2025people, bourke2020leaving}. While these perspectives establish allyship as dynamic and relational, they primarily treat allyship as an individual stance or as part of wider struggles for equity, rather than as something teams practice together. Our study extends this knowledge into the domain of collaborative virtual work in mixed-ability teams, showing how allyship emerges as a teamwork skill. Specifically, allyship supports productivity, fosters coordination within teams, and improves understanding of work processes. 

Our findings also show that allyship is not simply an individual endeavor but is shaped by team culture. Non-disabled participants emphasized that they learned allyship not through formal training, but through ongoing, informal processes such as observing others, receiving feedback, and adapting to evolving accessibility norms. These processes were deeply influenced by the broader team environment. In teams that fostered psychological safety, normalized discussions of accessibility, and framed access as a shared responsibility, non-disabled participants felt supported in experimenting with new behaviors, reflecting on mistakes, and sustaining inclusive practices. This insight complements prior work on collaborative learning by emphasizing that allyship learning is not only individual but also contextually scaffolded. Psychologically safe teams can function as learning environments \cite{edmondson1999psychological} that can enable the development of inclusive practices. 

Moreover, allyship was not unidirectional from non-disabled team members toward disabled team members. Disabled team members actively shaped the allyship process by initiating conversations, clarifying preferences, and offering guidance. This highlights allyship as a co-constructed process, extending the concept of relational access \cite{bennett_interdependence_2018} and emphasizing the agency of disabled team members in cultivating inclusive team norms.

\update{\textbf{Recommendation for Team Practice.} Teams can strengthen allyship by supporting it as an ongoing learning process through reflection and practice. For example, team members can regularly reflect on what kinds of support worked well, what caused discomfort, and how team members can do better next time. Regular reflections, such as debriefing after a meeting or adding an “access check-in” at the end of a project, can help keep these reflections consistent. Non-disabled members can learn how their actions affect others and disabled members can share what support feels appropriate or effective. When these conversations are normalized, mistakes can become opportunities for learning. Over time, this process can help allyship become a shared learning process.} 

\update{\textbf{Design Opportunity.}} 
\update{Virtual collaboration tools can help teams carry out regular reflections about allyship and learn allyship skills.}

\update{Existing platforms like Zoom, Microsoft Teams, and Slack already shape user behavior through nudges such as unmute reminders \cite{zoom_muting_unmuting}, hand-raise notifications \cite{teams-raise-hand-support}, or follow-up integrations \cite{followupbot-slack}. These mechanisms could be extended to foster allyship as a developmental practice.} 

\update{For example, systems could integrate anonymous “access check-in” prompts after meetings to ask about what support felt helpful or unhelpful. This supports the team practice of reflecting on how allyship was received, helping team members understand the impact of their actions and adjust over time.}

\update{Platforms could also offer microlearning nudges for inclusive behaviors, such as offering guidance on how to extend help without overstepping, or prompting follow-up with team members after meetings. These nudges can support the reflective, adaptive learning process of allyship. Importantly, disabled team members should be able to customize or review these nudges, ensuring that suggested practices align with how they wish to be supported.}
\subsubsection{Creating Shared Expectations for Allyship}
\update{\textbf{Challenge.}} Our findings show that allyship developed not only through ongoing learning but also through the establishment of shared expectations that provided clarity for non-disabled team members. Prior work has noted how non-disabled participants often take initiative to support accessibility in collaborative settings \cite{shinohara_shadow_2011, yildiz_virtual_2023, mack_mixed_2021}, but we found that such proactive support could become problematic when it lacked communication about what kinds of help were needed or wanted (see Section \ref{allyship-overstep}). Effective allyship in mixed-ability teams was not simply a matter of taking initiative; it required negotiation and respect for the autonomy of disabled team members. Participants emphasized that unsolicited help, even when well-intentioned, was sometimes experienced as intrusive. Non-disabled team members instead learned to ask before acting, clarify rather than assume, and wait for cues rather than intervening preemptively.

These findings align with disability etiquette guidance \cite{ny-disability-etiquette} and principles from critical disability studies, which emphasize agency, interdependence, and the situated expertise of disabled individuals \cite{kittay2011ethics, carework-piepzna-samarasinha-2018}. We extend this work by showing how these values were actively negotiated and shaped in virtual teamwork. Disabled team members shaped allyship by clarifying preferences, setting boundaries, and offering guidance, while allies gained confidence through clearer expectations and reduced uncertainty.

\update{\textbf{Recommendation for Team Practice.} Teams should establish shared expectations for when and how to offer help. Instead of relying on assumptions or guessing what others need, team members can talk openly about preferences for support, such as when it is helpful to step in. Disabled team members might share their preferences, such as “I appreciate offers to help with describing images, but please ask first,” and non-disabled members can practice checking in before intervening. These conversations can happen at the start of collaboration, when new team members join, or when needs change over time. By setting these expectations together, team members can avoid uncertainty and provide support that is desired by disabled team members.}

\update{\textbf{Design Opportunity.}} 
\update{Virtual collaboration tools can support the establishment of shared expectations by making it easier for team members to communicate their accessibility-related preferences.}

\update{Existing features already provide mechanisms for users to communicate preferences or additional information. Zoom and Microsoft Teams, for example, include reaction icons \cite{zoom-nonverbal-feedback} and status indicators (e.g., “away,” “do not disturb”) \cite{teams-user-presence}, while Slack enables customizable statuses \cite{slack-status-availability}.}

\update{These features can be extended to help team members voluntarily indicate accessibility-related preferences after they are established with their team. For example, platforms could allow users to specify preferences in their profiles, such as “ask before offering help,” “visual descriptions when sharing screens are appreciated.” Making these preferences visible can support team members in remembering established expectations.}

\update{Tools could also prompt teams to revisit and have conversations about accessibility preferences over time, such as when team needs change or when new members join, to ensure that preferences are up to date. This can help teams maintain shared expectations over time.}
\subsection{Limitations and Future Work} 
\update{Below, we outline limitations of our work and future research directions that can further advance understanding of mixed-ability teamwork in virtual collaboration settings.} 

The community would benefit from future work that explores a broader scope of participants. \update{Our participant sample included more disabled than non-disabled people. Future research should aim to include more non-disabled participants to provide more perspectives on mixed-ability collaboration. Expanding this representation would help reveal additional challenges and practices from non-disabled perspectives.}

Additionally, future studies with a wider range of disabilities could help researchers better understand the different needs and challenges in mixed-ability teams. This approach would enrich the understanding of how different disabilities interact within team settings and how diverse accessibility practices impact teamwork.

\update{While our semi-structured interview protocol was designed to elicit concrete examples of instances or interactions, interviews rely on memories of past experiences, which may be selective or shaped by participants’ interpretations over time. Future studies using observational or longitudinal approaches, such as contextual inquiry, could capture more immediate, situated accounts and complement the semi-structured interview method.} Observational studies would allow for a direct, real-time examination of teamwork and accessibility practices, while longitudinal studies could offer insights into how accessibility practices and teamwork evolve, such as the ways mixed-ability teams adapt to changing needs and challenges.

Furthermore, teamwork encompasses a vast array of dimensions and dynamics, many of which remain underexplored in mixed-ability teams. Future research could examine dimensions such as leadership. \update{Future work could also explore how people in mixed-ability teams define teamwork dimensions and compare with non-mixed-ability teams to examine whether participants define teamwork dimensions differently.} Exploring these additional aspects could provide a richer understanding of how mixed-ability teams collaborate and highlight further opportunities to design tools to support their unique needs. 
\section{Conclusion}
In this paper, we conducted semi-structured interviews with 18 participants, 12 with and 6 without disabilities, in mixed-ability teams that collaborate in virtual settings. We found that accessibility practices were not merely accommodations for disabled team members, but mechanisms that shaped teamwork for both disabled and non-disabled team members. Accessibility practices influenced how teams built rapport, coordinated responsibilities, and navigated accountability challenges. Non-disabled participants also described learning allyship through ongoing collaboration with their disabled team members. \update{Building on these insights, we recommend team practices for stronger teamwork and design opportunities for virtual collaboration tools to support those team practices.} Overall, our work points to a novel avenue where accessibility practices serve as a foundation for strong teamwork in virtual collaboration settings.
\appendix

%% file: participants.tex
\begin{table*}
  \caption{Participant demographics.}
  \label{tab:participant-demographics}
  \small
  \setlength{\tabcolsep}{4pt}
  \renewcommand{\arraystretch}{1.2}
  \centering
  \begin{tabular}{@{}l c c p{0.20\linewidth} p{0.15\linewidth} p{0.40\linewidth}@{}}
    \toprule
    \textbf{ID} & \textbf{Age} & \textbf{Gender} & \textbf{Disability} & \textbf{Field of Mixed-Ability Team} & \textbf{Demographics of Mixed-Ability Team} \\
    \midrule
    N1  & 21 & Female & N/A & Education & Physical disability; people without disabilities \\
    D2  & 30 & Female & Legally blind; physical disability & Technology & Physical disability; people without disabilities \\
    N3  & 19 & Female & N/A & Education & Physical disability; blind/low vision; people without disabilities \\
    D4  & 43 & Female & Totally blind; ADHD & Technology & Blind/low vision; ADHD; people without disabilities \\
    D5  & 26 & Female & ADHD & Technology & ADHD; Autism; people without disabilities \\
    D6  & 26 & Female & ADHD; Autism & Education & Autoimmune disease; ADHD; people without disabilities \\
    D7  & 46 & Female & Physical disability & Nonprofit & Physical disability; people without disabilities \\
    N8  & 24 & Male   & N/A & Nonprofit & Physical disability; people without disabilities \\
    D9  & 27 & Female & Chronic illness; neurodivergent & Education & Unspecified disabilities; people without disabilities \\
    D10 & 56 & Female & Physical disability & Nonprofit & Physical disability; people without disabilities \\
    D11 & 29 & Female & ADHD & Research & ADHD; chronic illness; people without disabilities \\
    N12 & 45 & Male   & N/A & Technology & Blind/low vision; people without disabilities \\
    D13 & 34 & Female & Physical disability; ADHD; Autism & Research & Autism; people without disabilities \\
    D14 & 38 & Female & Legally blind & Technology & Blind/low vision; d/Deaf and hard-of-hearing; physical disabilities; ADHD; dyslexia; people without disabilities \\
    D15 & 28 & Male   & Legally blind; ADHD; Autism & Technology & Blind/low vision; ADHD; Autism; people without disabilities \\
    N16 & 21 & Female & N/A & Research & d/Deaf and hard-of-hearing; people without disabilities \\
    D17 & 36 & Male   & ASD \& PTSD & Technology & People without disabilities \\
    N18 & 37 & Female & N/A & Nonprofit & Blind/low vision; people without disabilities \\
    \bottomrule
  \end{tabular}
\end{table*}

%% file: design-rec.tex
\begin{table*}[t]
  \caption{For each teamwork challenge identified in our Findings (Column~1), we list a recommendation for a team practice that can address the challenge (Column~2), and then outline opportunities for tools to be designed  to support the team practice (Column~3).}
  \label{tab:design-rec}
  \centering
  \begin{tabular}{p{0.25\linewidth} p{0.30\linewidth} p{0.38\linewidth}}
    \toprule
    \textbf{Teamwork Challenge} &
    \textbf{Recommendation for Team Practice to Address Challenge} &
    \textbf{Design Opportunity to Support Team Practice} \\
    \midrule

    Inaccessible tools lead to uneven task distribution and emotional strain for non-disabled team members (see Section~\ref{inaccessible-tools}). &
    Teams should openly discuss how accessibility barriers shift workload, review who took on extra responsibilities, and redistribute tasks collaboratively. &
    Tools should support teams in surfacing reassigned work and prompt reflections about workloads and frustrations (e.g., flag access barriers, highlight reassigned tasks, reflection reminders, mood check-ins). \\

    \addlinespace
    Accommodations without clear expectations create accountability challenges, frustration, and uncertainty (see Section~\ref{accountability}). &
    Teams should co-create explicit norms that clarify how flexibility will be offered, how long it will last, what follow-up is expected, and when check-ins will occur. &
    Tools should support teams in making expectations around flexibility and accountability visible and easy to revisit (e.g., flexibility tags and timers, prompts to renegotiate deadlines). \\

    \addlinespace
    Allyship is learned over time, but there is a lack of support for this learning process (see Section~\ref{allyship-learning}). &
    Teams should regularly reflect on what support worked, what caused discomfort, and how to improve. &
    Tools should support teams in carrying out regular reflections about allyship and learning allyship skills (e.g., anonymized accessibility reflection prompts, microlearning nudges for inclusive behaviors). \\

    \addlinespace
    Unclear or unsolicited support can undermine the autonomy of disabled team members (see Section~\ref{allyship-overstep}). &
    Teams should establish shared expectations for when and how help should be offered, ensuring support is given with consent and aligned with individual preferences. &
    Tools should support teams in communicating accessibility-related preferences and maintaining them over time (e.g., accessibility preference profiles, prompts to revisit preferences). \\

    \bottomrule
  \end{tabular}
\end{table*}